\documentclass[12pt]{article}

\begin{document}
\makeatletter
\@addtoreset{equation}{section}
\makeatother
\renewcommand{\theequation}{\thesection.\arabic{equation}}
\baselineskip 15pt 

\title{\bf A General Argument Against the Universal Validity of
the Superposition Principle}

\author{Angelo Bassi\footnote{e-mail: bassi@ts.infn.it}\\
{\small Department of Theoretical Physics of the University of Trieste,}\\
and \\
\\ GianCarlo Ghirardi\footnote{e-mail: ghirardi@ts.infn.it}\\
{\small Department of Theoretical Physics of the University of Trieste, and}\\
{\small the Abdus Salam International Centre for Theoretical Physics,
Trieste, Italy.}}

\date{}

\maketitle

\begin{abstract}

We reconsider a well known problem of quantum theory, i.e. the so called
measurement (or macro-objectification) problem, and we rederive the fact
that it gives rise to serious problems of interpretation. The novelty of our
approach derives from the fact that the relevant conclusion is
obtained in a completely general way, in particular, without
resorting to any of the assumptions of ideality which are
usually done for the measurement process.   The generality and
unescapability of our
assumptions (we take into account possible malfunctionings of
the apparatus, its unavoidable entanglement with the
environmment, its high but not absolute reliability, its
fundamentally uncontrollable features)  allow to draw the
conclusion that the very possibility of performing measurements
on a microsystem combined with the assumed general validity of
the linear nature of quantum evolution leads to a fundamental
contradiction.

\end{abstract}

\section{Introduction}

In most of the treatises on the foundations of Quantum Mechanics, when
discussing the measurement problem and its possible solutions, the
authors  make use of the von Neumann scheme for an
{\it ideal measurement process} \cite{vn}
since, due to its simplicity, it allows to grasp immediately
why the standard solution to such a problem is not satisfactory.
The von Neumann argument goes as follows.

Consider a microscopic system $S$ and one of its observables $O$.
Let $o_{n}$ be its eigenvalues (we  suppose the
spectrum of $O$ to be purely discrete and
non--degenerate), and $|o_{n}\rangle$ the corresponding eigenvectors.

Let us call $M$ the apparatus devised to measure the observable $O$ of
the microsystem $S$. One supposes that $M$ has a ready--state
$|M_{0}\rangle$, i.e. a state in which the apparatus is ready
to measure the considered property, plus a set of {\it mutually
orthogonal} states  $|M_{n}\rangle$ (and orthogonal to the
ready--state), which correspond to {\it different macroscopic
configurations} of the instrument, like different positions of a
pointer along a scale.

Finally, one assumes that the interaction between the microsystem $S$
and the apparatus $M$ is {\it linear} (since the
Schr\"odinger equation is supposed to govern all natural processes) and
that it yields a {\it perfect  correlation} between the initial state of
$S$ and the final state of  the apparatus, i.e.
\begin{equation} \label{eq0}
\makebox{Initial state: $\;\;|o_{n}\rangle\otimes|M_{0}\rangle$}\qquad
\longrightarrow\qquad
\makebox{Final state: $\;\;|o_{n}\rangle\otimes|M_{n}\rangle$};
\end{equation}
in this way one is sure that, if the final state of the apparatus is
$|M_{n}\rangle$ (i.e. the pointer, for example, is in the $n$--th
position along the scale), he can say that the state of the particle is
$|o_{n}\rangle$, and that the observable $O$ has the value $o_{n}$.

Within such a context, the measurement problem arises when the initial
state of the particle,  previous to the measurement, is not just one of
the vectors
$|o_{n}\rangle$ like in eq. (\ref{eq0}),
but a superposition of them, for example:
\[
|m + l\rangle\quad =\quad \frac{1}{\sqrt{2}}[ |o_{m}\rangle +
|o_{l}\rangle].
\]
In this case, due to the {\it linearity} of the quantum evolution, the
final state of the microsystem+apparatus is given by:
\begin{eqnarray} \label{eq01}
|m + l\rangle\otimes|M_{0}\rangle & = & \frac{1}{\sqrt{2}}\;[
|o_{m}\rangle + |o_{l}\rangle]\otimes|M_{0}\rangle \longrightarrow
\nonumber \\
& \longrightarrow & \frac{1}{\sqrt{2}}\,[|o_{m}\rangle\otimes
|M_{m}\rangle + |o_{l}\rangle\otimes|M_{l}\rangle ].
\end{eqnarray}
This final state is an entangled state of the microscopic system and
of the apparatus, and it is well known that (if one assumes that the
theory is complete, i.e., that the wave--function  contains {\it all} the
information about the system) in the considered case it is not {\it even
in principle}  legitimate to state that the properties associated to the
states
$|M_{m}\rangle$  or $|M_{l}\rangle$
are possessed by the apparatus (the same holds true for
the microsystem): as a consequence, the apparatus is not in any
macroscopic definite configuration. This is the essence of the quantum
measurement  problem.

The {\it standard} way out from this difficulty is given by the
so--called  {\it wave--packet reduction postulate}, which states that
``at the end of the {\it measurement} process'' the final vector in
(\ref{eq01}) reduces to one of its terms
\[
|o_{m}\rangle\otimes|M_{m}\rangle \qquad \makebox{or} \qquad
|o_{l}\rangle\otimes|M_{l}\rangle,
\]
with a probability given by the square modulus of the coefficient
associated to that term ($1/2$ for both outcomes, in our example).

It is obvious that the postulate of wave--packet reduction contradicts
the general validity of the Schr\"odinger equation and consequently it
does not represent
a satisfactory solution to the measurement problem. Accordingly,
many other (more or less
satisfactory) solutions have been proposed \cite{grw, bohm, wig, mm,
dh}.  In spite of this, various people \cite{zu, pri, lib} have suggested
that the  problem does not derive from the structure of quantum mechanics
(in  particular from the linear character of the quantum evolution),
or from the wave--packet reduction postulate, but from adopting
the over--simplified model of the measurement process put forward by
von Neumann. If one  takes into account a more realistic model, they
argue, the measurement problem would turn into a false one, and
there would be no need to modify the interpretation of Quantum Mechanics,
or  to put forward a new theory.

In particular, the following assumptions have been criticized:
\begin{itemize}
\item That the measuring apparatus can be prepared in a
precise state $|M_{0}\rangle$: since the instrument is a macroscopic
object with many degrees of freedom, it is impossible to know its
precise state at any given time.

\item That one can safely neglect the interactions between the apparatus
and the surrounding environment. In fact, the interaction with the
environment (which is referred to as decoherence)  produces
essentially a randomization
of the phases associated to the different components of the
wave--function, a process which can be seen as an {\it apparent} collapse
of the  wave--function into one of these components.

\item  That the final states of the apparatus, corresponding to
perceptively different macroscopic configurations of the apparatus
itself, are orthogonal: actually,
different states usually correspond to different {\it positions} of some
component of the instrument, and since no wave--function can have
compact support in  configuration space (because of the quantum
evolution),  wave--functions corresponding to different states
cannot, in general, be orthogonal.

\item That the final state of the apparatus gets perfectly
correlated to the initial state of the microscopic system:
this is an highly idealized characteristic which is not shared by any
realistic physical instrument.

\end{itemize}

Just to give some examples of the above situation, we consider it
appropriate to mention explicitly  some statements by distinguished
physicists, supporting the idea that the measurement
problem is a false one:
\begin{enumerate}
 \item H. Primas, during a conference in Finland \cite{pri},
has stated:
\begin{quote} {\small
... the measurement problem of quantum mechanics as it is discussed by
theoreticians and philosophers is an extremely ill posed problem ...
Measurements of the first kind are unrealistic and completely
irrelevant for experimental science ... Replace in all textbooks and
philosophical treatises on quantum mechanics the word `measurement' by
the proper expression `measurement of the first kind' and add a
footnote: measurements of the first kind are idealizations which
never play any role in experimental science.}
\end{quote}
In this article we will prove rigorously that also when consideration
is given to  very reasonable and realistic schemes of measurement
processes, the  problem cannot be avoided. As a
consequence, Primas' statements are not pertinent.

 \item In an article appeared in Physics Today \cite{zu}, W.
Zurek claims that, because of the interaction with the environment, the
\begin{quote} {\small
... coherent superposition of ... states [like those of eq.
(\ref{eq01})] ... is continuously reduced to
a mixture. A preferred basis of the detector, sometimes called a
``pointer basis'', has been singled out ... {\it Decoherence prevents
superpositions of the preferred basis states from persisting}.}
\end{quote}
Despite what Zurek claims, it is well known that decoherence, {\it by
itself}, does not ``prevent superpositions of the preferred basis
states from persisting'', and the theorem we will prove in this paper
will show that Zurek's conclusions are incorrect.

 \item In a quite recent book [9], some of the authors (whose conceptual
position with respect
 decoherence is slightly different from the one of refs. [7,8]) seem to
consider
 decoherence a satisfactory explanation of why ordinary macroscopic objects
appear to
behave classically. For example, in Section 8.3 where decoherence and
spontaneous
localization models [2] are compared, we find it written:
\begin{quote} {\small
... stochastic collapse models may be an interesting concept for the
purpose of illustrating and discussing the measurement problem in
quantum theory, but there seems to be no phenomenological necessity
for its introduction, neither in connection with the appearance of
classical properties ({\it for which environmental decoherence already
provides a satisfactory explanation}), nor prompted by new phenomena.}
\end{quote}
We stress again that decoherence does not represent by itself a solution to the
macro--objectification problem in general, and to the measurement
problem in particular: only if one adopts a
precise interpretation (like bohmian mechanics, decoherent histories, ...),
decoherence
can be of help in solving these problems.  Unfortunately, all the above
authors forget
to inform us of the  interpretation they adopt (which, of course, cannot be
the
standard one, since it includes  the postulate of wave--packet
reduction which, as we have said, is contradictory).
\end{enumerate}

As we have already mentioned, in this paper  we will
consider a completely general and realistic model
for the measurement process, and we will show that
superpositions of different macroscopic
configurations of  macro--objects cannot be avoided within a strict
quantum mechanical scheme. Correspondingly, the appearence of
macroscopic situations which are incompatible with our definite
perceptions about the objects of our experience is inescapable. This
``empasse" can only   be eliminated either
by  adopting a precise and unambiguous interpretation which differs
from the  orthodox one, or by modifying the theory itself\footnote{An
explicit  proof that releasing the request of an ideal measurement
does not allow to  circumvent the measurement problem  can be found
in the well  known book by d'Espagnat
\cite{de}; however, his proof is much more complex and much  less
general than the one we are going to present here.}.

\section{The Microscopic System}

Let us start our general discussion of the measurement process by
analyzing the microscopic system whose properties we want to measure;
we consider for simplicity the simplest system upon which non--trivial
measurements can be performed, i.e. a system ($S$) whose associated
Hilbert space (${\mathcal H}_{S}$) is two--dimensional  --- like the one
describing the spin of an electron, or the polarization states of a
photon --- and we consider  an observable
$O$  having two
different eigenvalues; let us call
$|\makebox{u}\rangle$ and $|\makebox{d}\rangle$ the eigenstates
associated to these eigenvalues. For definiteness, we will consider a single
such system and we will call ``spin'' its degree of freedom;  we will say that
the particle has ``spin Up'' when it is in state $|\makebox{u}\rangle$, and
that it has ``spin Down'' when it is in state $|\makebox{d}\rangle$.
Besides these two states, also their superpositions can be
taken into account, like for example:
\[
| \makebox{u + d}\rangle \quad =\quad \frac{1}{\sqrt{2}}\,\,[\,
|\makebox{u}\rangle\, +\, |\makebox{d}\rangle\,],
\]
a vector describing a new state, ``spin Up + spin Down'', of the
particle. Without any loss of generality,
we will assume that, by resorting to appropriate procedures,
one can ``prepare'' the system $S$ in any one of the above considered
states.

We remark that we could have considered more general
physical systems, like compound ones, and observables having a more
complicated spectrum (e.g. the continuos spectrum of position).
Anyway, in accordance with the generally accepted position  that
microsystems can be prepared in  a precise quantum state and with the
(nowadays) common experimental practice  to  handle
single particles and to measure their (discrete) spin (or  polarization,
in the case of photons) states, we have chosen to work with  very simple
microsystems as the one we are considering here. Moreover, we also assume
that, after the preparation,  the system is in a precise and known
state, and that  it can be treated as isolated from the rest of the
world, at least untill  the measurement process begins\footnote{In
mathematical terms,  we assume that, {\it prior to the measurement
process},  the wave--function of the universe
factorizes into the wave--function of the particle times the
wave--function of the rest of the world.}. We stress that if one denies
these assumptions it is not clear what he takes quantum theory to be
about.

\section{The Measuring Apparatus}

A measuring apparatus is a {\it macroscopic} system which,
interacting with the microsystem whose properties one is interested in
ascertaining, ends up into a state more or less correlated with the
eigenstates of the observable it is devised to measure. The different
possible outcomes of the measurement are supposed to be correlated to
{\it perceptively different macroscopic configurations} of a part of
the apparatus, e.g. different positions
of the pointer (for analogic instruments), different numbers on a
display (for digital ones), different spots on a photographic plate,
different plots on a screen, and so on.
For simplicity, in what follows we will
assume that the apparatus has a pointer  movable along a scale, whose
position registers the result of the measurement.

Contrary to microsystems, the measuring apparatus,
being a macroscopic object, has many degrees
of freedom, most of which ---  in particular the microscopic ones
--- we cannot control at all; and of the macroscopic ones,
like the position of the pointer, we can have only a very limited
control. Moreover,
the apparatus, due to its dimensions, is always interacting with the
environment (whose degrees of freedom are also essentially out of
control). Following this line of reasoning, one can remark
that the apparatus --- or at least its constituents --- existed quite a
long before the measurement is performed, so it had all the time to
interact (even if only weakly) with a large part of the universe, or
perhaps with all of it.
All these interactions make, to a large extent, unknown and
uncontrollable the state of the macroscopic systems which enters into
play. In spite of this difficulty, in order to keep our analysis as
general as possible, we will take them all into account.

According to the above discussion, we should, in general, speak of
different situations of the ``whole universe'', even though our
``reading'' refers only to the degrees of freedom of the  pointer; accordingly,
we shall indicate the statevectors we will deal with in the following way:
\[
        |A\; \alpha\rangle.
\]
These vectors belong to the Hilbert space  associated to the
apparatus,  the environment, and in the most general case to the
whole universe.
$A$ is a label which indicates that the pointer of the apparatus is in a
specific macroscopic configuration, i.e. one which we perceive and we identify
with a  specific position along the scale. In first  approximation, we
could say
that
$A$ is essentially the value $x$  characterizing the ``projection operator''
$|x\rangle\langle x|$ ($|x\rangle$ being an ``improper''
statevector of the Hilbert space of the pointer)
giving the exact  position of (e.g., the centre of mass of) the pointer
along the scale. But it is evident that no system  can be prepared in
such a state (since it is impossible to measure a  continuous variable
with a perfect accuracy); and even if it were possible to do so, the
hamiltonian evolution would immediately change that state: thus the
pointer cannot ever be in an eigenstate of
$|x\rangle\langle x|$.

We could try to improve our model by taking into account not precise
positions along the scale, but small intervals $\Delta(x) = [ x -
\delta, x +
\delta ]$, and claiming that ``the pointer is at position $x$'' when the
wave--function is an eigenstate of the projection operator which
projects onto the interval\footnote{Of course, here we are
considering for simplicity a one--dimensional situation; the argument
can be easily generalized to the three--dimensional case.} $\Delta(x)$ of the
position of the centre of mass. If one takes such a position, the label $A$
characterizing our general state
$|A\; \alpha\rangle$  refers to any wave--function having  such a property, of
course with the interval
$\Delta(x)$ replaced by the interval $\Delta(A)$: as a
consequence, for the considered state
we can  claim that ``the pointer is at position A''. However, also this
approach
is  not viable since the hamiltonian evolution transforms any
wave--function with compact support into a wave--function with a
non--compact one; this fact gives rise to what has sometimes been called the
``tail problem'', a problem which cannot be avoided,
and which  renders rather
delicate the task of making precise the idea of ``an object being
somewhere'' within a quantum mechanical framework.

In the light of the above discussion, we consider a very
general physical situation: we will call $V_{A}$ the set of all
(normalized) vectors $ |A\; \alpha\rangle$ for which we are {\it
allowed} to say that ``the pointer of the apparatus is at position
$A$'' or, stated differently, that ``the universe is in a
configuration which we perceive as one corresponding to the statement:
the pointer is at $A$''. We do not put any restriction to the vectors
belonging to $V_{A}$: they can represent wave--functions with or
without tails, more or less localized in space, and so on; we do not even
resort to projection operators to characterize these states. The only
physical requirement we put forward is that, if the pointer admits two
macroscopically different positions along the scale (let us call them
$A$ and $B$), then any two vectors corresponding to such different
configurations must be ``almost orthogonal''. This requirement can be
translated into the following mathematical relation: denoting as
$V_{B}$ the set of all normalized vectors corresponding to the
statement ``the pointer is at $B$'' while $V_{A}$, as before, contains
all the vectors corresponding to
the statement ``the pointer is at $A$'', we must have:
\begin{equation} \label{req}
\inf_{\begin{array}{l}
\makebox{\footnotesize $|A\;\alpha\rangle \,\in\, V_{A}$} \\
\makebox{\footnotesize $|B\;\beta\rangle \,\in\, V_{B}$}
\end{array}}\, \| |A\;\alpha\rangle -
|B\;\beta\rangle \|\, \geq\, \sqrt{2} - \eta \qquad\qquad \eta \ll 1,
\end{equation}
i.e. the minimum distance between the vectors of the two above sets
cannot differ too much from $\sqrt{2}$, which is the distance
between two orthogonal normalized states. We recall that the orthogonality
request of the standard measurement theory is done to be sure to be dealing
with
strictly mutually exclusive situations. Obviously such a request can be
partially released (as we are doing here) but not given up completely if one
wants to be able to ``read'' the outcome in a fundamentally non ambiguous way.
It is evident that  (\ref{req}) is a necessary requirement if
one pretends that {\it different} macroscopic
positions of the pointer (and of any other system) represent {\it mutually
exclusive} configurations of the object\footnote{Obviously, here we are making
reference to a genuinely quantum description (with the completeness
assumption).
In alternative interpretations or formulations of the theory, orthogonality is
not necessary  to guarantee macroscopic differences. Tipically, in hidden
variables theories one could have non orthogonal wave--functions and different
values for the hidden variables such that the associated physical
situations are
macroscopically different.}.

Let us now comment on the second parameter
$\alpha$ characterizing our states: this is an index which takes into account
all other degrees of freedom that are out of control\footnote{From
the mathematical point of view, $\alpha$ stands for the
eigenvalues of a  complete set of commuting observables for the whole
universe, exception made for the position of the pointer.}; thus, two
vectors labeled by $A$, but with different values for $\alpha$, refer to
the ``same'' macroscopic configuration for the pointer (or, in general, of the
``part of the universe we perceive"), while they describe two different states
for the rest of the universe (e.g., given a certain atom of the pointer,
it might be in the ground state when the
state is $|A\; \alpha\rangle$, while it might be in an excited state
when it is $|A\;\beta\rangle$).

Since the microscopic particle has two spin--states, if we want to use
the apparatus to distinguish them we have to assume that the pointer
admits at least two
macroscopically different positions ($U$ and $D$) along the
scale\footnote{The idea is that, if we perform
the measurement and we find the pointer in the position labeled by
$U$, then we can claim the ``the result of the measurement is that the
spin of the particle is Up''; similarly, if we find
the pointer in $D$, then we can say that ``the spin of the particle
is Down''.}; the previous argument  requires to assume that
there exist two sets $V_{U}$ and $V_{D}$: the first one  contains all
the vectors corresponding to the situation in which the ponter can be
said to point at ``U'', while the second set contains all those
vectors associated to the statement ``the pointer is at D''.
Moreover, these two sets must be
almost orthogonal in the sense of (\ref{req}):
\begin{equation} \label{req2}
\inf_{\begin{array}{l}
\makebox{\footnotesize $|U\;\alpha\rangle \,\in\, V_{U}$} \\
\makebox{\footnotesize $|D\;\beta\rangle \,\in\, V_{D}$}
\end{array}}\, \| |U\;\alpha\rangle -
|D\;\beta\rangle \|\, \geq\, \sqrt{2} - \eta \qquad\qquad \eta \ll 1,
\end{equation}
One interesting property of $V_{U}$ and $V_{D}$ (which is shared by
any couple of sets satisiyng (\ref{req2})) is that they have no vectors
in common: in fact, it is easy to see that if $V_{U}$ and $V_{D}$ had
such a common vector, then the minimum distance between them would be
zero, a fact which would contradict (\ref{req2}). From the physical
point of view, this property is obvious
since a vector belonging both to $V_{U}$ and to $V_{D}$ would be a vector
for which we could claim both that ``the pointer points at U'' and that ``the
pointer points at D'', a contradictory situation since ``U'' and ``D''
correspond to  macroscopically different situations.

\section{The Preparation of the Apparatus}

A measuring instrument must be prepared before one performs a
measurement, i.e. one has to arrange the apparatus in such a way that it is
ready to interact with the microscopic system and give a result;
following the discussion of the previous section, it is evident that
the initial state--vector must carry an index $\alpha$ which takes
into account the state of the rest of the universe:
accordingly, we will denote the initial state--vector as
$|A_{0}\;\alpha\rangle$, where $A_{0}$ indicates that the pointer
``is'' in the ready ($A_{0}$) state.

Anyway, we remember that, besides the measuring instrument, we have
also to prepare the microsystem in a precise state,
and moreover we have assumed that after the preparation and
immediately before the measurement process, the microsystem itself is
isolated from the rest of the universe;
the initial state--vector for the whole universe
can then be written as:
\[
|A_{0}\;\alpha\rangle \quad =\quad
|\makebox{spin}\rangle\otimes|A_{0}\;\overline{\alpha}\rangle,
\]
where
$\overline{\alpha}$ specifies the state of the whole universe, with
the exception of the initial state of the micro--particle and the
initial ``position" of the pointer; $|\makebox{spin}\rangle$ is the
initial state--vector of the particle.

Obviously, also in the process of preparing the apparatus we cannot control
the state of
the universe so that  we do not  know the precise initial state
$|A_{0}\;\overline{\alpha}\rangle$:  in fact, in any specific situation any
value for
the index
$\overline{\alpha}$ will occur with a given probability
$p(\overline{\alpha})$, which in general is unknown to us --- but, of
course, it has to satisfy appropriate requirements  we will
discuss in what follows. Accordingly,
the initial setup, for the apparatus and the microscopic particle,
will be described as follows:
\[ \makebox{Initial Setup} \quad =\quad \left\{\,
|\makebox{spin}\rangle\otimes|A_{0}\;\overline{\alpha}\rangle,
\;\;\; p(\overline{\alpha}) \right\},
\]
where $p(\overline{\alpha})$ gives the probability distribution
of the remaining, uncontrollable, degrees of freedom.

\section{The Measurement Process}

If one assumes that Quantum Mechanics governs all physical systems,
the measurement process, being an interaction between two
quantum systems, is governed by a unitary operator $U(t_{I}, t_{F})$.
Suppose the initial state of the microsystem is $|\makebox{u}\rangle$ and
the one of the apparatus (plus the rest of the universe)
is $|A_{0}\; \overline{\alpha}\rangle$; then, during the
measurement, the whole universe evolves in the following way:
\begin{equation} \label{evu}
|\makebox{u}\rangle\otimes|\makebox{$A_{0}$}\;
\overline{\alpha} \rangle \quad
\longrightarrow
\quad U(t_{I}, t_{F})\left[\, |\makebox{u}\rangle\otimes|
\makebox{$A_{0}$}\; \overline{\alpha} \rangle
\right]\quad =\quad |\makebox{F u}\; \overline{\alpha} \rangle,
\end{equation}
while, if the initial state of the microsystem is
$|\makebox{d}\rangle$,  one has:
\begin{equation} \label{evd}
|\makebox{d}\rangle\otimes|\makebox{$A_{0}$}\;
\overline{\alpha} \rangle \quad
\longrightarrow
\quad U(t_{I}, t_{F})\left[\, |\makebox{d}\rangle\otimes|
\makebox{$A_{0}$}\; \overline{\alpha} \rangle
\right]\quad =\quad |\makebox{F d}\; \overline{\alpha} \rangle.
\end{equation}
Some comments are needed.
\begin{itemize}
\item Note that in the above equations (\ref{evu}) and (\ref{evd})
the index $\overline{\alpha}$ distinguishes various
possible and uncontrollable situations of the measuring apparatus in its
``ready'' state. Once the initial state is fully specified also the final one,
being the evolution unitary, is perfectly and unambiguously determined.
Accordingly, such a state is appropriately characterized by the same index
$\overline{\alpha}$. Note also that, while the state $|\makebox{$A_{0}$}\;
\overline{\alpha}
\rangle$ belongs to the Hilbert space of the whole universe exception
made for the micro--particle, the state $|\makebox{F d}\;
\overline{\alpha}\rangle$ now includes also the particle.

        \item Contrary to what one does in the ideal--measurement
scheme of von Neumann, we do not assume that  the final state is
factorized; thus, in general
\[ |\makebox{F u}\; \overline{\alpha} \rangle \quad \neq \quad
|\makebox{u}\rangle \otimes |A_{U}\; \overline{\alpha}\rangle.
\]
\item In particular, we do not suppose that the final state of the
microsystem be the same as the initial one: we allow the measurement
process to modify in a significant way the state of the particle;
it could even destroy the particle.
\end{itemize}

        The only thing we require is that {\it the measuring apparatus
is reliable to a high degree}, i.e. that it can safely be used to measure the
state of the microsystem since, in most cases, it gives the correct answer.
This means that if the initial state of the microsystem (prior to the
measurement) is $|\makebox{u}\rangle$, then the final state
$|\makebox{F u}\; \overline{\alpha}\rangle$ must belong {\it in most of the
cases} to $V_{U}$, while, if the initial state of the particle is
$|\makebox{d}\rangle$, then the final state $|\makebox{F d}\;
\overline{\alpha}\rangle$ must {\it almost always} belong to $V_{D}$.
Note that by not requiring full reliability, we take into account also the
possibility that the measuring  instrument  gives the wrong results, though
pretending that such mistakes occur quite seldom.

        It is possible to formalize the above reliability requests in the
following way.
Let us consider the set $K$ of all subsets $J$ of the possible values
that the index $\overline{\alpha}$ can assume and let us equip it with the
following (natural) measure:
\[
\mu(J)\quad =\quad \sum_{\overline{\alpha}\in J} p(\overline{\alpha}).
\]
Let us also define the two following sets:
\[
\begin{array}{lcl}
J^{-}_{U} & = & \left\{\overline{\alpha}\;\; \makebox{such that:} \;\;
|\makebox{F u}\; \overline{\alpha}\rangle \not\in V_{U} \right\},
\end{array}
\]
\[
\begin{array}{lcl}
J^{-}_{D} & = & \left\{\overline{\alpha}\;\; \makebox{such that:} \;\;
|\makebox{F d}\;\overline{\alpha}\rangle \not\in V_{D} \right\}.
\end{array}
\]
$J^{-}_{U}$ is the sets of all the indices $\overline{\alpha}$ such that
the states $|\makebox{F u}\;\overline{\alpha}\rangle$ do not
correspond to the outcome {\it ``the pointer is at position U''}, despite the
fact that prior to the measurement the state of the particle was
$|\makebox{u} \rangle$.
Similarly, $J^{-}_{D}$ corresponds to the states $|\makebox{F d}\;
\overline{\alpha} \rangle$ for which we cannot claim that {\it
``the pointer is in D''}, even if the initial state of the system was
$|\makebox{d}\rangle$.
Let also $J^{+}_{U} = {\mathcal C}J^{-}_{U}$ be the
complement set of $J^{-}_{U}$, and $J^{+}_{D} = {\mathcal C}
J^{-}_{D}$ the complement of $J^{-}_{D}$.

Given this, the requirement that the instrument is reliable can be
mathematically expressed in the following way:
\begin{equation} \label{req3}
\mu(J^{-}_{U})\;\; \leq\;\; \epsilon \qquad\quad
\mu(J^{-}_{D})\;\; \leq\;\; \epsilon, \qquad\quad \epsilon \ll 1,
\end{equation}
i.e. the probability that the final position of the pointer does not
match the initial spin--value of the particle is very small, this
smallness being controlled by an appropriate parameter $\epsilon$
expressing the efficiency of the measuring device and which, as such, can
change (always remaining very small) with the different actual measurement
procedures one can devise.

Of course, it is easy to derive also limits on the
measure for the complements of the above sets:
\[
\mu(J^{+}_{U}) \;\; \geq \;\; 1 - \epsilon \qquad\qquad
\mu(J^{+}_{D}) \;\; \geq \;\; 1 - \epsilon.
\]
We need to take into account also the two sets:
$J^{-} = J^{-}_{U} \cup J^{-}_{D}$ and
$J^{+} = {\mathcal C}J^{-} = J^{+}_{U} \cap J^{+}_{D}$; they satisfy
the following relations:
\[
\mu(J^{-}) \;\; \leq \;\; 2\epsilon \qquad\qquad
\mu(J^{+}) \;\; \geq \;\; 1 - 2\epsilon.
\]
Again, all these limits simply state that, being the apparatus
reliable, the probability that --- at the end of the measurement
process --- the pointer is not in the correct position is very
small, {\it if} the initial state of the particle was either
$|\makebox{u}\rangle$ or $|\makebox{d}\rangle$.

        It is useful to remark that, having taken into account
the possibility that the measuring instrument can make
mistakes, we can easily include also the possibility that it
fails to interact at all with the microsystem, thus giving no
result: in such a case, the pointer remains in the
``ready--state'', and the corresponding vector belongs to the
set $J^{-}$. In fact, let us consider the set $V_{0}$
associated to the ``ready--state'', as we did for the two  sets
$V_{U}$ and $V_{D}$ referring to the ``U'' and ``D''  positions
of the pointer. By the same argument as before,  $V_{0}$ is
disjoint from the two sets $V_{U}$ and $V_{D}$, since the
``ready--state'' is macroscopically different from the ``U''
and ``D'' states; consequently, if the vector, at the end of
the measuring process, belongs to
$V_{0}$, it cannot belong either to $V_{U}$ or to $V_{D}$.

        We have mentioned the possibility that the apparatus misses to
detect the
particle because such an occurrence affects (and in some case in an
appreciable way) many experimental situations (e.g. the efficiency of
photodetectors is usually quite low). This does not pose any problem to
our treatment: in fact, we can easily circumvent this difficulty by simply
disregarding (just as it is common practice in actual experiments) all cases
in which a detector should register something but it doesn't. The previous
analysis and the sets we have identified by precise mathematical criteria
must then be read as referring exclusively to the cases in which the
apparatuses register an outcome.

\section{Proof of the Theorem}

        In the previous section we have reconsidered the measurement
 scheme, first analyzed by von Neumann, and we have reformulated it
 on very general grounds; in fact, the only requirements at the basis
of our discussion are the following two:
\begin{enumerate}
\item That the quantum evolution of any physical system is {\it linear},
since it is governed by the Schr\"odinger equation;
\item That any two sets, like $V_{U}$ and $V_{D}$, containing vectors
corresponding to {\it different macroscopic configuration} of a
macro--object are almost orthogonal:
\begin{equation} \label{req4}
\inf_{\begin{array}{l}
\makebox{\footnotesize $|U\;\alpha\rangle \,\in\, V_{U}$} \\
\makebox{\footnotesize $|D\;\beta\rangle \,\in\, V_{D}$}
\end{array}}\, \| |U\;\alpha\rangle -
|D\;\beta\rangle \|\, \geq\, \sqrt{2} - \eta \qquad\qquad \eta \ll 1.
\end{equation}
\end{enumerate}
We think that everybody would agree that {\it any real} measurement
situation (if it has to be described in  quantum mechanical terms),
shares these two properties\footnote{As already remarked, request 2) can be
violated in hidden variables theories. On the other hand, request 1)
is purposedly violated in dynamical reduction theories. Since both theories
account for the objectification of macroscopic properties, they must
necessarily violate one of the two requests.}.

Starting with these very simple premises we can now easily  show that
quantum mechanics must face the problem of the occurrence of
superpositions of macroscopically different states of the
apparatus, and in general of a macro--system.

In our terms, the ``measurement problem'' arises (as usual)
when the initial spin--state of the
particle is not $|\makebox{u}\rangle$ or $|\makebox{d}\rangle$, as we
have considered in the previous sections, but a superposition of them,
like the state $|\makebox{u + d}\rangle$ of section 2, which can be easily
prepared in a
laboratory. In such a case, due to the linearity of the evolution, the
final state of
the particle+apparatus system will be:
\[
\begin{array}{lcl}
|\makebox{u + d}\rangle\otimes|\makebox{$A_{0}$}\;\overline{\alpha}\rangle
\quad & \longrightarrow &
\quad U(t_{I}, t_{F})\left[\, |\makebox{u + d}\rangle\otimes|
\makebox{$A_{0}$}\;\overline{\alpha}\rangle
 \right]\quad = \quad |\makebox{F u+d}\;\overline{\alpha}
\rangle\\
& & \quad= {\displaystyle
\frac{1}{\sqrt{2}}\left[ |\makebox{F u}\;\overline{\alpha}\rangle +
|\makebox{F d}\;\overline{\alpha}\rangle \right]}.
\end{array}
\]
It is now very simple to prove that for each $\overline{\alpha}$
belonging to $J^{+}$, $|\makebox{F u+d}\;\overline{\alpha}\rangle$
{\it cannot} belong either to $V_{U}$ or to $V_{D}$. In fact, let us
suppose that it belongs to $V_{U}$ (the proof in the case in which
it is assumed to belong to $V_{D}$ is analogous); since the distance
between $|\makebox{F u+d}\;\overline{\alpha}\rangle$  and
$|\makebox{F d}\;\overline{\alpha}\rangle$ is:
\begin{eqnarray*}
\| |\makebox{F u+d}\;\overline{\alpha}\rangle -
|\makebox{F d}\;\overline{\alpha}\rangle \| & = &
\| 1/\sqrt{2}\, |\makebox{F u}\;\overline{\alpha}\rangle
+ \left( 1/\sqrt{2} - 1\right) \,
|\makebox{F d}\;\overline{\alpha}\rangle \| \quad \leq \quad \\
& \leq & \frac{1}{\sqrt{2}} + 1 - \frac{1}{\sqrt{2}}
\quad = \quad 1,
\end{eqnarray*}
we get a contradiction, because
$|\makebox{F u+d}\;\overline{\alpha}\rangle$
belongs to $V_{U}$ and $|\makebox{F d}\;\overline{\alpha}\rangle$
belongs to $V_{D}$, and relation (\ref{req4}) must hold between any two
vectors of these two sets: this proves that
$|\makebox{F u+d}\;\overline{\alpha}\rangle$  cannot belong either
to $V_{U}$ or to $V_{D}$. Of course, by the same argument,
we can also prove that, for all $\overline{\alpha}\,\in\,J^{+}$,
the index of the apparatus cannot be in any
other macroscopic position different from ``U'' and ``D''.

We have shown that, for all
$\overline{\alpha}\,\in\,J^{+}$ and for all measurements processes
in which the apparatus registers an outcome, the vector
$|\makebox{F u+d}\;\overline{\alpha}\rangle$ {\it does not allow to
assign  any macroscopic definite position to the
index  of the apparatus, not even  one different from ``U'' or ``D''}.
Stated differently, the large majority of the initial apparatus states, when
they are triggered by the superposition $|\makebox{u + d}\rangle$, end up in a
state which does not correspond to any definite position (in our geneal
language
of section 3, to any definite situation of the part of the universe we
perceive), i.e. one paralleling our definite perceptions.

We believe that our formulation represents  the most general
proof of the unavoidability of  the macro-objectification problem: to have a
consistent picture one must accept that in a way or another the linear nature
of the dynamics must be broken.
We are fully aware that such a conclusion is an old one
which is largely shared by those interested
in the foundational aspects of quantum mechanics. But we also think that
our derivation of this result is useful and interesting for the
absolutely minimal (and physically unavoidable) requests on which it is
based, i.e. that  one can prepare microscopic systems in precise states
which are eigenstates of a quantum  observable and that when this is done
and the considered observable is measured, one can  get reliable
information about the eigenvalue of the observable itself, by
appropriate  amplification procedures leading to perceivably different
macroscopic situations of the  universe.

\end{document}